\documentclass[12pt]{iopart}
\pdfoutput=1
\usepackage{color}
\usepackage{graphicx}
\usepackage{amsmath, amsthm, amssymb}
\usepackage[normalem]{ulem}
\definecolor{Gray}{gray}{0.7}
\usepackage{citesort}
\usepackage{tabularx}
\usepackage{amsfonts}
\usepackage{mathtools}
\usepackage{tikz}
\usetikzlibrary{positioning,decorations.pathmorphing,decorations.markings,arrows}
\usepackage{physics}
\usepackage[colorlinks=true,linkcolor=blue,citecolor=blue]{hyperref}
\usepackage{latexsym,amsfonts,color,amsthm}
\usepackage{enumerate}
\usepackage{fancyhdr} 
\usepackage[square, numbers, comma, sort&compress]{natbib}
\usepackage{glossaries}
\usepackage{colortbl}
\usepackage[export]{adjustbox}
\usepackage[titletoc]{appendix}

\newcommand{\be}{\begin{equation}}
\newcommand{\ee}{\end{equation}}
\newcommand{\bea}{\begin{eqnarray}}
\newcommand{\eea}{\end{eqnarray}}

\begin{document}

\title[Mpemba effect in anisotropically driven inelastic Maxwell gases]{Mpemba effect in anisotropically driven inelastic Maxwell gases} 
\author{Apurba Biswas$^{1,2}$, V. V. Prasad$^{3}$, and
R. Rajesh$^{1,2}$}

\address{$^1$ The Institute of Mathematical Sciences, C.I.T. Campus, Taramani, Chennai 600113, India}
 \address{$^2$ Homi Bhabha National Institute, Training School Complex, Anushakti Nagar, Mumbai 400094, India}
\address{$^3$ Department of Physics, Cochin University of Science and Technology,  Kochi 682022, India}
 \ead{apurbab@imsc.res.in, prasad.vv@cusat.ac.in, rrajesh@imsc.res.in}

\date{\today}

\begin{abstract}
Through an exact analysis, we show the existence of Mpemba effect in an anisotropically driven inelastic Maxwell gas, a simplified model for granular gases, in two dimensions. Mpemba effect refers to the couterintuitive phenomenon of a hotter system relaxing to the steady state faster than a cooler system, when both are quenched to the same lower temperature. The Mpemba effect has been illustrated in earlier studies on  isotropically driven granular gases, but its existence requires non-stationary initial states, limiting experimental realisation. In this paper, we demonstrate the existence of the Mpemba effect in anisotropically driven granular gases even when the initial states are non-equilibrium steady states. The precise conditions for the Mpemba effect, its inverse, and the stronger version, where the hotter system cools exponentially faster are derived.
\end{abstract}

\maketitle

\section{\label{sec1-Introduction}Introduction}
In recent times, there has been considerable interest in the Mpemba effect, a counterintuitive phenomenon wherein a hot system equilibrates faster than a cooler system when quenched to a low temperature. It was initially predicted for water~\cite{1952meteorologica,Mpemba_1969}. Though many reasons have been attributed to the cause of the Mpemba effect in water including convection~\cite{vynnycky-convection:2015}, evaporation~\cite{Mirabedin-evporation-2017}, dissolved gases~\cite{katz2009hot},  supercooling~\cite{david-super-cooling-1995}, hydrogen bonding~\cite{zhang-hydrbond1-2014,tao-hydrogen-2017,Molecular_Dynamics_jin2015mechanisms} and non-equipartition of energy~\cite{gijon2019paths}, the precise cause is still debated. One such study has even cast doubts about the existence of the Mpemba effect in water~\cite{NoMpemba}. Though the Mpemba effect, as described in the case of water, involves a phase transition where the final phase is ice and the initial phase water or steam, similar Mpemba effect  has been observed in other physical systems that does not involve a phase transition. The other physical systems where this effect has been demonstrated experimentally includes  clathrate hydrates~\cite{paper:hydrates}, magnetic alloys~\cite{chaddah2010overtaking}, polylactides~\cite{Polylactide} and more recently in colloidal systems~\cite{kumar2020exponentially,kumar2021anomalous}. 

Analysis of model-based systems also shows the existence of the Mpemba effect in spin systems~\cite{PhysRevLett.124.060602,Klich-2019,klich2018solution,das2021should}, three state  Markovian systems~\cite{Lu-raz:2017}, spin glasses~\cite{SpinGlassMpemba}, molecular gases in contact with a thermal reservoir~\cite{moleculargas,gonzalez2020mpemba,gonzalez2020anomalous} and  granular systems~\cite{Lasanta-mpemba-1-2017,Torrente-rough-2019,mompo2020memory,PhysRevE.102.012906,biswas2021mpemba}. For the case of spin systems~\cite{PhysRevLett.124.060602,Klich-2019,klich2018solution,das2021should} and three state Markovian systems~\cite{Lu-raz:2017}, the initial probability distributions describing the hot and the cold systems correspond to their equilibrium (Boltzmann) distribution and then they are evolved to the equilibrium of the final cold temperature following the Markovian dynamics. The exact condition for the existence of the Mpemba effect is derived by analysing the distance between the probability distributions during the relaxation process. Moreover, such systems also show the existence of the inverse Mpemba effect~\cite{Lu-raz:2017} where an initially colder system can heat up faster than an initially warmer system, the strong Mpemba effect~\cite{Klich-2019} where certain initial states lead to an exponentially faster cooling and also exhibit optimal heating protocols~\cite{PhysRevLett.124.060602} in which precooling leads to faster heating. For the case of spin glasses, two systems are prepared which are in contact with different temperature thermal baths. The time evolution of their energy density (instantaneous energy per spin) is analysed to demonstrate the Mpemba effect when both the  systems are quenched using a cold temperature thermal bath. For the systems of molecular gases (elastic collisions) in contact with a thermal bath~\cite{moleculargas,gonzalez2020mpemba}, the Mpemba effect is analysed using the mean kinetic energy of the constituent molecules. For the case of molecular gas of single species, in contact with a thermal bath~\cite{moleculargas}, the Mpemba effect is due to the coupling of mean kinetic energy with the excess kurtosis of the velocity distribution function in the presence of non-linear viscous drag. On the other hand, for binary mixture of molecular gases~\cite{gonzalez2020mpemba,gonzalez2020anomalous} in contact with a background fluid, the Mpemba effect is due to the coupling of mean kinetic energies of the individual components of the binary gas. In both cases for molecular gases, the Mpemba effect is illustrated only for non-stationary initial states.
 
In this paper, we focus on granular gases, a dilute composition of driven inelastic particles. Granular gases are of special interest as it is one possible area where a strong interplay between experiment and theoretical analysis for an interacting particle system is possible. At the same time, it is also an example of a system that is far from equilibrium. The Mpemba effect has been demonstrated  in driven granular gases in  few different contexts.  To study the Mpemba effect in granular systems, two systems are prepared at two different granular temperatures (mean kinetic energy of particles). On quenching to a lower temperature (by changing the driving parameters), the Mpemba effect is said to be present if the temperatures of the two systems cross each other while relaxing to the final stationary state. For a system of smooth monodispersed particles~\cite{Lasanta-mpemba-1-2017}, the Mpemba effect is achieved by the coupling of mean kinetic energy with the excess kurtosis  of the velocity distribution function. An exact analysis was possible for the case of an inelastic Maxwell gas, wherein the rate of collision is simplified to be independent of relative velocity, and it was shown that there has to be non-trivial correlations among the initial velocities of the particles to achieve the Mpemba effect~\cite{PhysRevE.102.012906}. The Mpemba effect was also demonstrated for a system of rough granular gas~\cite{Torrente-rough-2019},  granular gas of viscoelastic particles~\cite{mompo2020memory} and in inertial suspensions of granular particles~\cite{PhysRevE.103.032901}. In all these analysis, the initial states of the systems are non-stationary for the Mpemba effect to be achieved. This is a drawback, as achieving special non-stationary states in experiments is much harder than attractive stationary states.

To achieve the Mpemba effect in a granular system with stationary initial conditions, a couple of systems have been put forward. Through an exact analysis of a driven 
binary granular Maxwell gases~\cite{PhysRevE.102.012906}, it was shown that the coupling between the mean kinetic energies of the two components of the binary gas leads to the Mpemba effect,  the inverse Mpemba effect and the strong Mpemba effect starting from steady state initial conditions. Here, a mechanism of driving the two types of particles differently is required, which may be difficult to achieve in practice. For a monodispersed gas in two dimensions, it was recently shown that it is possible to achieve the Mpemba effect, its inverse and the strong counterpart with initial stationary states provided the driving is anisotropic (different in the two directions)~\cite{biswas2021mpemba}. This was established based on an analysis of the Enskog-Boltzmann equation for driven granular gases with the simplifying assumption that the velocity distribution is a gaussian. By linearising the theory about the stationary states, it is shown that the Mpemba effect can be achieved by simply tuning the driving strengths, thus making it an effective system for experimental realisation of the effect. Results from event-driven simulations are consistent with the results from the linearised theory~\cite{biswas2021mpemba}.

In this paper, we do an exact analysis of  the system of monodispersed inelastic gas with anisotropic driving based on the inelastic Maxwell model in two dimensions. Compared to the system studied in Ref.~\cite{biswas2021mpemba} where the rate of collision is proportional to the relative velocity, in the Maxwell gas, the rate of collision is independent of the relative velocity. While this makes the Maxwell gas more unrealistic, it renders it more amenable to exact analysis, at the same time retaining the qualitative features. This advantageous feature has been exploited in obtaining more rigorous results in both freely cooling granular gas~\cite{PhysRevE.66.011309,Baldassarri_2002,Ernst_2002,Krapivsky_2002} as well as in the velocity distributions of driven granular gases~\cite{PhysRevE.65.040301,PhysRevE.66.062301,PhysRevE.68.011305,prasad2014high,Prasad:14,prasad2017velocity,binary_maxwell}. The equations for the time evolution of the relevant two point velocity correlations for the Maxwell gas form a closed set of equations~\cite{Prasad:19}. We analyse these equations to determine the condition and the parameter regime for the existence of the Mpemba effect. With our exact analysis of the anisotropically driven Maxwell gas, we are able to put the results of Ref.~\cite{biswas2021mpemba}, which depended on many simplifying assumptions, on a more sound footing. We show that the Mpemba and the inverse Mpemba effects exist for steady state initial conditions which can be prepared by tuning the physical parameters defining the system. In this analysis, we also demonstrate the existence of the strong Mpemba effect where for certain specific initial steady states, the equilibration rate is exponentially faster compared to any other initial steady states. 
 
 The remainder of the paper is organised as follows. Section~\ref{sec2-The Model} contains the definition of the model. In Sec.~\ref{sec3-Two point correlations}, we show that the time evolution of two point velocity-velocity correlations do not depend on higher order correlations and form a closed set of equations. This allows for an exact calculation of the steady state mean kinetic energies. In Sec.~\ref{sec4-The Mpemba effect}, we define the Mpemba effect and demonstrate its existence for an anisotropically driven granular gas. In Sec.~\ref{sec5-Limiting case where the driving is along one direction}, we discuss the case where driving is limited to only one direction. Section~\ref{sec6-Summary and discussion} contains the summary of results and discussion of their various implications.

\section{\label{sec2-The Model} The Model}
Consider a monodispersed granular gas composed of $N$ identical particles. We label the particles by $i=1, \ldots, N$ and denote their two dimensional velocities by $\emph{\textbf{v}}_\emph{\textbf{i}}=(v_{ix},v_{iy})$. These velocities evolve in time through momentum conserving inelastic binary collisions and external driving. A pair of particles $i$ and $j$ collide at a rate $\lambda_c /N$. The factor $1/N$ in the collision rates ensures that the total rate of collisions between $N[N-1]/2$ pairs of particles are proportional 
to the system size $N$. The new velocities $\emph{\textbf{v}}'_\emph{\textbf{i}}$ and $\emph{\textbf{v}}'_\emph{\textbf{j}}$ are given by 
\bea
\begin{split}
\boldsymbol{v}'_i=\boldsymbol{v}_i -\alpha[(\boldsymbol{v}_i-\boldsymbol{v}_j).\boldsymbol{\hat{e}}]\boldsymbol{\hat{e}},  \\
\boldsymbol{v}'_j=\boldsymbol{v}_j +\alpha[(\boldsymbol{v}_i-\boldsymbol{v}_j).\boldsymbol{\hat{e}}]\boldsymbol{\hat{e}},
\end{split} \label{eq:collision}
\eea
where
\bea
\alpha=\frac{1+r}{2},
\eea
$r$ being the co-efficient of restitution, and $\boldsymbol{\hat{e}}$ is the unit vector along the line joining the centres of the particles at contact. We assume that $\boldsymbol{\hat{e}}$ takes a value uniformly from $[0, 2 \pi)$ for each collision.
In addition to collisions, the system evolves through external driving. Each particle is driven at a rate $\lambda_d$. During a driving event, the new velocity $\emph{\textbf{v}}'_\emph{\textbf{i}}$ is given by
\begin{eqnarray}
v'_{ix}=-r_{wx}v_{ix}+\eta_x, \quad -1<r_{wx} \leq 1, \nonumber \\
v'_{iy}=-r_{wy}v_{iy}+\eta_y, \quad -1<r_{wy} \leq 1, \label{eq:driving}
\end{eqnarray}
where $r_{wx}$ and $r_{wy}$ are parameters  of driving and $\boldsymbol{\eta}$ is a noise taken from a fixed distribution $\Phi(\boldsymbol{\eta})$. We denote the second moment of the noise distribution by $\sigma^2_{x}$ and $\sigma^2_{y}$:
\be
\sigma^2_{k}=\int^{\infty}_{-\infty} d\eta_k  \eta^2_k \Phi( \boldsymbol{\eta}),\quad k=x,y. \label{eq:variance}
\ee
Note that $\sigma^2_{x} \neq \sigma^2_{y}$ or $r_{wx}\neq r_{wy}$ corresponds to anisotropic driving and will introduce an anisotropy in the resultant  velocity distribution of the particles. Such a driving scheme [Eq.~(\ref{eq:driving})] leads the system to a steady state and has been used extensively in earlier studies~\cite{prasad2014high,Prasad:14,prasad2017velocity}. The physical motivations for the form of driving may be found in Refs.~\cite{Prasad:18,Prasad:19}, where positive $r_w$'s can be identified as the coefficient of restitution of collisions between particle and a vibrating wall.

The model has two simplifying features. The spatial degrees of freedom have been neglected. This corresponds to the well-mixed limit where the spatial correlations between particles are ignored. In addition, the collision rates are independent of the relative velocity of the colliding particles. This corresponds to the so called Maxwell limit. 

Let $P(\boldsymbol{v},t)$ denote the probability that a randomly chosen particle has velocity $\boldsymbol{v}$ at time $t$. Its time evolution is given by
\begin{align}
\frac{d}{dt}P(\boldsymbol{v},t)&=\lambda_c \int \int \int d\boldsymbol{\hat{e}} d\boldsymbol{v}_1 d\boldsymbol{v}_2 P(\boldsymbol{v}_1,t) P(\boldsymbol{v}_2,t) \delta(\boldsymbol{v}_1- \alpha[(\boldsymbol{v}_1-\boldsymbol{v}_2).\boldsymbol{\hat{e}}]\boldsymbol{\hat{e}}- \boldsymbol{v}) \nonumber \\
& +\lambda_d \int \int d\boldsymbol{\eta} d\boldsymbol{v}_1 \Phi(\boldsymbol{\eta}) P(\boldsymbol{v}_1,t)\delta[-r_w\boldsymbol{v}_1+ \boldsymbol{\eta}-\boldsymbol{v} ]  -\lambda_c  P(\boldsymbol{v},t) -\lambda_d P(\boldsymbol{v},t), \label{eq:time ev} 
\end{align}
where the first and third terms on the right hand side describe the gain and loss terms due to collisions while the second and fourth terms describe the gain and loss terms due to driving.
\section{\label{sec3-Two point correlations}Two point correlations}
 We are interested in the time evolution of the following two-point correlation functions:
\begin{eqnarray}
E_x(t)=\frac{1}{N}\sum^{N}_{i=1} \langle v^2_{ix}(t) \rangle,&C_{x}(t)=\frac{1}{N (N-1)}\sum^{N}_{i=1} \sum_{\substack{j=1 \\ j \neq i}}^{N} \langle v_{ix}(t) v_{jx}(t) \rangle, \nonumber \\
E_y(t)=\frac{1}{N}\sum^{N}_{i=1} \langle v^2_{iy}(t) \rangle,& C_{y}(t)=\frac{1}{N (N-1)}\sum^{N}_{i=1} \sum_{\substack{j=1 \\ j \neq i}}^{N} \langle v_{iy}(t) v_{jy}(t) \rangle, \nonumber \\
E_{xy}(t)=\frac{1}{N }\sum^{N}_{i=1} \langle v_{ix}(t) v_{iy}(t) \rangle,& C_{xy}(t)=\frac{1}{N (N-1)}\sum^{N}_{i=1} \sum_{\substack{j=1 \\ j \neq i}}^{N} \langle v_{ix}(t) v_{jy}(t) \rangle. \label{2 point correlations} 
\end{eqnarray} 
$E_x(t)$ and $E_y(t)$ denote the mean kinetic energies of the particles along $x$ and $y$ directions respectively. $E_{xy}(t)$ denote the correlations between $v_x$ and $v_y$ of the same particle whereas $C_x(t)$, $C_y(t)$ and $C_{xy}(t)$ denote the velocity-velocity correlations between pairs of particles. The time evolution for these correlation functions can be obtained starting from Eq.~(\ref{eq:time ev})~\cite{prasad2014high,Prasad:14,prasad2017velocity,Prasad:19,PhysRevE.102.012906}. These may be written compactly in a matrix form as 
\begin{align}
\frac{d\boldsymbol{\tilde{\Sigma}}(t)}{dt} =\boldsymbol{\tilde{R}}  \boldsymbol{\tilde{\Sigma}}(t) + \boldsymbol{\tilde{D}},
\label{two-point-correlation-matrix-evolution}
\end{align}
where the column vectors $\boldsymbol{\tilde{\Sigma}}(t)$ and $\boldsymbol{\tilde{D}}$ are given by:
\begin{eqnarray}
\boldsymbol{\tilde{\Sigma}}(t)=[E_x(t),E_y(t),E_{xy}(t),C_x(t),C_y(t),C_{xy}(t)]^T,  \\
\boldsymbol{\tilde{D}}=[\lambda_d \sigma^2_x,\lambda_d \sigma^2_y, 0, 0, 0, 0]^T.
\end{eqnarray}
While the matrix $\boldsymbol{\tilde{R}}$ can be written for any $N$, in the thermodynamic limit $N\rightarrow \infty$, it simplifies to
 
\begin{equation}
\boldsymbol{\tilde{R}}\!=\!\left[\begin{array}{cccccc} 
A_{1}+A^{xx}_4 &A_2 &0 & -A_1 &-A_2 & 0  \\
A_2 & A_{1}+A^{yy}_4 &0 & -A_2 & -A_1 &0 \\
0 & 0 & -A_3+A^{xy}_4 &0 &0 &A_3 \\
0 & 0 & 0 & 2A^x_5& 0 &0 \\
0 & 0 & 0 &0 & 2A^y_5 & 0 \\
0 & 0 & 0 & 0 & 0 & A^x_5+A^y_5
\end{array}\right].
\label{eq:Rmatrix}
\end{equation} 
 The constants $A_1, A_2, A_3, A^{ij}_4, A^{i}_5$ are given by:
 \bea
 \begin{split}
 &A_1=\frac{3}{4}\lambda_c \alpha^4-\lambda_c \alpha,\quad  A_2=\frac{\lambda_c \alpha^4}{4}, \\
 &A_3=\lambda_c \alpha(1-\frac{\alpha}{2}),\quad A^{ij}_4=-\lambda_d(1-r_{wi}r_{wj}),\\
 &A^i_5=-\lambda_d(1+r_{wi}), \quad \textnormal{where}~ i,j \in (x,y).
 \end{split}
 \eea
In the steady state, the left-hand side of Eq.~(\ref{two-point-correlation-matrix-evolution}) equals zero. After taking the thermodynamic limit ($N\rightarrow \infty$), we obtain the steady state values of the different two point correlation functions as
\begin{eqnarray}
E_x=\frac{\lambda_d \big[\big(4\lambda_d(1-r^2_{wy})+\lambda_c \alpha(4-3\alpha) \big)\sigma^2_x +\alpha^2 \lambda_c \sigma^2_y \big]}{\mathcal{F}}, \label{energy-x-ss} \\
E_y=\frac{\lambda_d \big[\big(4\lambda_d(1-r^2_{wx})+\lambda_c \alpha(4-3\alpha) \big)\sigma^2_y +\alpha^2 \lambda_c \sigma^2_x \big]}{\mathcal{F}}, \label{energy-y-ss} \\
E_{xy}=C_x=C_y=C_{xy}=0, \label{correlations-ss}
\end{eqnarray}
where
\begin{align}
\mathcal{F}=4\lambda_d^2(1-r^2_{wx})(1-r^2_{wy})+\alpha \lambda_c \lambda_d(4-3\alpha)(2-r^2_{wx}-r^2_{wy})+2\alpha^2\lambda^2_c (2-3\alpha+\alpha^2).
\end{align}
From the structure of $\boldsymbol{\tilde{R}}$~[see Eq.~(\ref{eq:Rmatrix})], it is evident that the time evolution of velocity-velocity correlations only depend (linearly) on other velocity-velocity correlations and do not depend on the mean kinetic energies. Thus, if in the initial state, these correlations are zero, then they remain zero for all times. Since we will be considering only initial states that are stationary, the velocity-velocity correlations are initially zero [see Eq.~(\ref{correlations-ss})] and will continue to remain zero for all times.
 
We therefore set these velocity correlations to zero and write the time evolution for only the non-zero quantities,  $E_x$ and $E_y$ as:
\begin{equation}
\frac{d\boldsymbol{\Sigma}(t)}{dt}=~\boldsymbol{R}\boldsymbol{\Sigma}(t)+\boldsymbol{S},\label{time-ev}
\end{equation}
where
\begin{align}
&\boldsymbol{\Sigma}(t)=\begin{bmatrix}
E_{x}(t), 
E_{y}(t)
\end{bmatrix}^T, \\
&\boldsymbol{S}=\begin{bmatrix}
\lambda_d \sigma^2_x, 
\lambda_d \sigma^2_y
\end{bmatrix}^T,
\end{align}
and  $\boldsymbol{R}$ is a  $2\times2$ matrix, whose entries are given by
\begin{align}
\begin{split}
R_{11}=&\frac{3}{4}\lambda_c \alpha^2 -\lambda_c \alpha -\lambda_d(1-r^2_{wx}) , \quad R_{12}=\frac{\lambda_c}{4}\alpha^2,\\ 
R_{22}=&\frac{3}{4}\lambda_c \alpha^2 -\lambda_c \alpha -\lambda_d(1-r^2_{wy}), \quad R_{21}=\frac{\lambda_c}{4}\alpha^2 . \\ 
\end{split}
\end{align}

It is convenient to work in a different set of variables than $E_x(t)$ and $E_y(t)$. We introduce the total energy, $E_{tot}$, and the difference in energies, $E_{dif}$, as:
\begin{eqnarray}
E_{tot}=E_x+E_y,
\label{E1 define}\\
E_{dif}=E_x-E_y.
\label{E2 define}
\end{eqnarray}
Note that since the driving is anisotropic, $E_{dif}\neq0$ in general.

The time evolution equations for $E_{tot}$ and $E_{dif}$ can be expressed, starting from Eq.~(\ref{time-ev}), as
\begin{equation}
\frac{d\boldsymbol{E}(t)}{dt}=-~\boldsymbol{\chi}\boldsymbol{E}(t)+\boldsymbol{D}, \label{time ev E}
\end{equation}
where
\begin{align}
&\boldsymbol{E}(t)=\begin{bmatrix}
E_{tot}(t), 
E_{dif}(t)
\end{bmatrix}^T, \\
&\boldsymbol{D}=\begin{bmatrix}
\lambda_d (\sigma^2_x+\sigma^2_y), 
\lambda_d (\sigma^2_x-\sigma^2_y)
\end{bmatrix}^T,
\end{align}
and $\boldsymbol{\chi}$ is a $2\times2$ matrix with the components of the matrix given by 
\begin{align}
\begin{split}
\chi_{11}&=\frac{2\lambda_c\alpha(1-\alpha)+\lambda_d(2-r^2_{wx}-r^2_{wy})}{2}, \quad \chi_{12}=\frac{\lambda_d(r^2_{wy}-r^2_{wx})}{2},\\
\chi_{22}&=\frac{\lambda_c\alpha(2-\alpha)+\lambda_d(2-r^2_{wx}-r^2_{wy})}{2}, \quad \chi_{21}=\frac{\lambda_d(r^2_{wy}-r^2_{wx})}{2}. \label{appendix : chi}
\end{split} 
\end{align}
Equation~(\ref{time ev E}) can be solved exactly by linear decomposition using the eigenvalues $\lambda_\pm$ of $\boldsymbol{\chi}$:
\begin{equation}
\lambda_\pm=\frac{1}{4}\Big[2 \lambda_d (2-r^2_{wx}-r^2_{wy}) + \alpha \lambda_c(4 -3 \alpha)\pm \sqrt{4 \lambda^2_d(r^2_{wy}-r^2_{wx})^2+ \alpha^4 \lambda^2_c}\Big].
\end{equation}

It is straightforward to show that $\lambda_{\pm}>0$ with $\lambda_+>\lambda_-$. The solution for $E_{tot}(t)$ and $E_{dif}(t)$ is
\begin{align}
\begin{split}
E_{tot}(t)-\langle E_{tot}\rangle&=K_+e^{-\lambda_+t}+ K_-e^{-\lambda_-t}, \\
E_{dif}(t)-\langle E_{dif}\rangle&=L_+e^{-\lambda_+t}+ L_-e^{-\lambda_-t},
\end{split} \label{sol E1 one driven}
\end{align}
where $\langle E_{tot}\rangle$ and $\langle E_{dif}\rangle$ are steady state values of $E_{tot}(t)$ and $E_{dif}(t)$ respectively. The coefficients $K_+, K_-, L_+$ and $L_-$ along with $\langle E_{tot}\rangle$ and $\langle E_{dif}\rangle$  are given in Eq.~(\ref{appendix coeff both components driven}). 
 These coefficients depend only on the system parameters and initial conditions. Equation~(\ref{sol E1 one driven}) gives the full time dependent solution for the energies, and we utilise them to demonstrate the Mpemba effect.

\section{\label{sec4-The Mpemba effect}The Mpemba effect in an anisotropically driven gas}
In this section, we show and determine the conditions for the existence of the Mpemba effect in the anisotropically driven monodispersed Maxwell gas based on the analysis of $E_{tot}(t)$ and $E_{dif}(t)$ [see Eq.~(\ref{sol E1 one driven})]. The Mpemba effect in granular systems has been defined as follows~\cite{Lasanta-mpemba-1-2017,Torrente-rough-2019,PhysRevE.102.012906,mompo2020memory,biswas2021mpemba}. Consider two systems $P$ and $Q$ which have identical  parameters except for the pair of driving strengths, $\sigma^2_x$ and $\sigma^2_y$. We will choose $E_{tot}$ of $P$ to be higher. Note that the systems $P$ and $Q$ are initially in steady states. We denote their steady state values for the energies by $[E^P_{tot}(0),E^P_{dif}(0)]$ and $[E^Q_{tot}(0),E^Q_{dif}(0)]$ respectively. Both the systems are then quenched to a common steady state having lower energy compared to the initial steady state energies of $P$ and $Q$. This is achieved by changing the driving strengths of $P$ and $Q$ to the common driving strengths, $\sigma^2_x$ and $\sigma^2_y$ of the final steady state, keeping all the other parameters of both the systems constant.

We say that the Mpemba effect is present if  the two trajectories $E^P_{tot}(t)$ and $E^Q_{tot}(t)$ cross each other at some finite time $t=\tau$ at which
\begin{equation}
E^P_{tot}(\tau)=E^Q_{tot}(\tau).
\end{equation}

To obtain the value of $\tau$, we equate the energies for $P$ and $Q$ from Eq.~(\ref{sol E1 one driven}) to obtain
\begin{equation}
K^P_+e^{-\lambda_+\tau}+ K^P_-e^{-\lambda_-\tau}=K^Q_+e^{-\lambda_+\tau}+ K^Q_-e^{-\lambda_-\tau}, \label{evolution eq at crossing}
\end{equation}
whose solution is
\begin{equation}
\tau=\frac{1}{\lambda_+-\lambda_-}\ln \Big[\frac{K^P_+-K^Q_+}{K^Q_--K^P_-} \Big].
\end{equation}

In terms of the parameters of the initial steady states, $\tau$ reduces to
\begin{equation}
\tau=\frac{1}{\lambda_+-\lambda_-} \ln \Big[\frac{\chi_{12}\Delta E_{dif}-(\lambda_--\chi_{11})\Delta E_{tot}}{\chi_{12}\Delta E_{dif}-(\lambda_+-\chi_{11})\Delta E_{tot}}  \Big], \label{crossing time one driven}
\end{equation}
where
\begin{align}
\begin{split}
&\Delta E_{tot}=E^P_{tot}(0)-E^Q_{tot}(0), \\
&\Delta E_{dif}=E^P_{dif}(0)-E^Q_{dif}(0).
\end{split} \label{eq: delta E}
\end{align}

For the Mpemba effect to be present, we require that $\tau >0$. Since $\lambda_+>\lambda_-$, the argument of logarithm in Eq.~(\ref{crossing time one driven}) should be greater than one. Simplifying, we obtain the criterion for the crossing of the two trajectories to be
\begin{equation}
\frac{\Delta E_{tot}}{\Delta E_{dif}}<\frac{2\lambda_d (r^2_{wy}-r^2_{wx})}{\lambda_c\alpha^2 + \sqrt{4 \lambda^2_d(r^2_{wy}-r^2_{wx})^2+ \alpha^4 \lambda^2_c}}. \label{condition one driven}
\end{equation}

\begin{figure}
\centering
\includegraphics[width=\columnwidth]{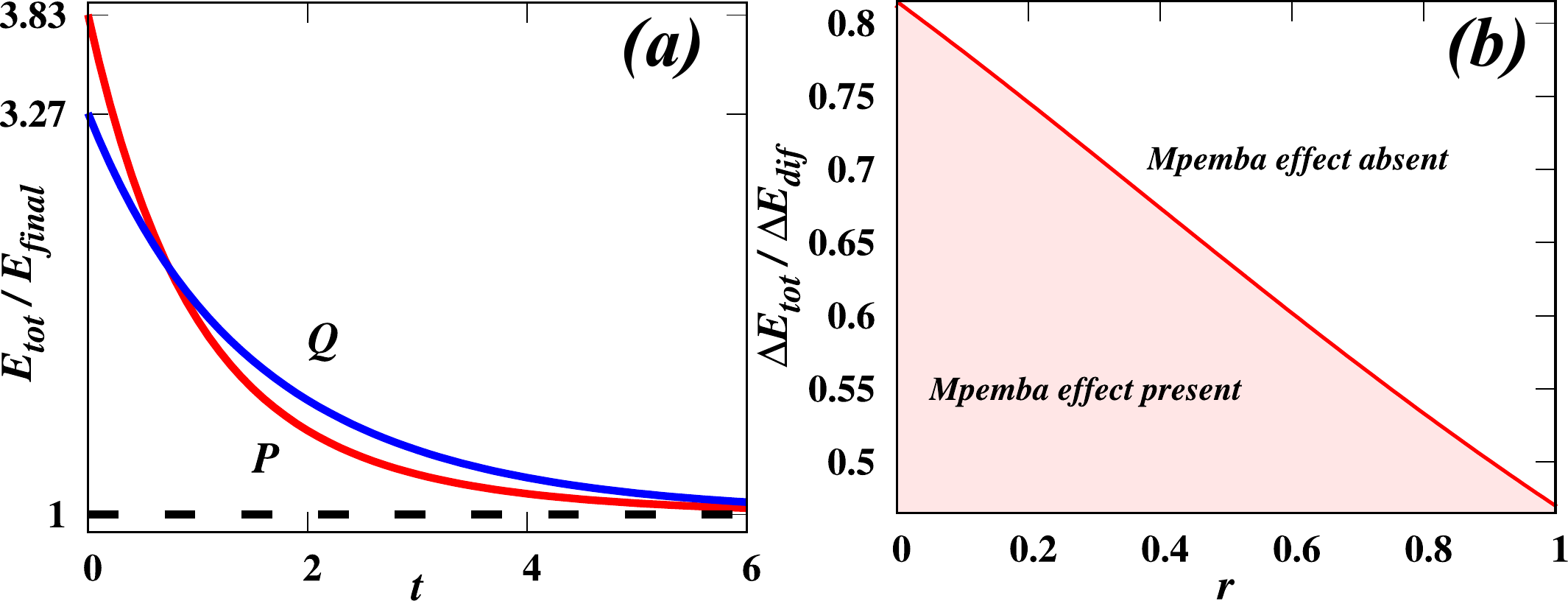}
\caption{(a) The time evolution of the total energy, $E_{tot}(t)$ for anisotropically driven systems $P$ and $Q$ of a two dimensional inelastic Maxwell gas, driven along both the directions of the plane,  with initial conditions $E^P_{tot}(0)$=20.27, $E^Q_{tot}(0)$=17.32, $E^P_{dif}(0)$=-7.93 and $E^Q_{dif}(0)$=6.26 such that $E^P_{tot}(0)>E^Q_{tot}(0)$, which satisfies the condition for the Mpemba effect as described in Eq.~(\ref{condition one driven}). The other parameters decribing the systems are chosen to be $r$=0.3, $r_{wx}=0.88$ and $r_{wy}=0.39$. $P$ relaxes to the steady state faster than $Q$, though its initial energy is larger. The time at which the trajectories cross each other is $\tau=0.73$ as given by Eq.~(\ref{crossing time one driven}). (b) $\Delta E_{tot}/\Delta E_{dif}$--$r$ phase diagram showing regions where the Mpemba effect is observed and $r$ is the coefficient of restitution. The line corresponds to a particular choice of the parameters $r_{wx}=0.2$, $r_{wy}=0.8$, $\lambda_c=1.0$ and $\lambda_d=1.0$. The region below the line given by Eq.~(\ref{condition one driven}) denotes the set of steady state initial conditions that show the Mpemba effect whereas the region on the other side of the line corresponds to initial states that do not show the Mpemba effect.}
\label{fig01}
\end{figure}

The right hand side of Eq.~(\ref{condition one driven}) depends only on the intrinsic parameters of the system and it is always less than one (since $\alpha, \lambda_c>0$). On the other hand, the ratio $\Delta E_{tot}/\Delta E_{dif}$,  depends on the initial steady state energies of $P$ and $Q$~[see Eq.~(\ref{eq: delta E})]. In the stationary state, the ratio $\Delta E_{tot}/\Delta E_{dif}$  is given by 
\begin{flalign}
\frac{\Delta E_{tot}}{\Delta E_{dif}}=\frac{
\big[\lambda_d (1-r^2_{wy}) + \alpha \lambda_c(2-\alpha)\big] \Delta \sigma^2_x 
+ \big[2\lambda_d (1-r^2_{wx}) + \alpha \lambda_c(2-\alpha)\big] \Delta \sigma^2_y
}{
2\big[\lambda_d (1-r^2_{wy}) + \alpha \lambda_c(1-\alpha)\big] \Delta \sigma^2_x
- \big[\lambda_d (1-r^2_{wx}) + \alpha \lambda_c(1-\alpha)\big] \Delta \sigma^2_y
},  \label{appendix: ss ratio Delta E}
\end{flalign}
where,
\bea
\Delta \sigma^2_i=(\sigma^P_i)^2-(\sigma^Q_i)^2,\quad i\in (x,y).
\eea
Equation~(\ref{appendix: ss ratio Delta E}) shows that the ratio $\Delta E_{tot}/\Delta E_{dif}$ depends on the intrinsic parameters of the system as well as the driving strengths, $\sigma^2_x$ and $\sigma^2_y$. As a result, the driving strengths can be appropriately tuned, keeping all the other intrinsic parameters identical for both $P$ and $Q$, to prepare the initial conditions that satisfy Eq.~(\ref{condition one driven}).
In Fig.~\ref{fig01}(a), we consider such a situation where Eq.~(\ref{condition one driven}) is satisfied. Here, the systems $P$ and $Q$ have identical intrinsic parameters but the pair of driving strengths, $\sigma^2_x$ and $\sigma^2_y$, are different for the two systems. The trajectories cross at the point as predicted by Eq.~(\ref{crossing time one driven}). It is clear that though $P$ has larger initial energy than $Q$, it relaxes faster compared to the latter.

Figure~\ref{fig01}(b) illustrates the phase space (initial conditions), based on Eq.~(\ref{condition one driven}), where the Mpemba effect is observable. In the figure, the line denotes the variation of right hand side of Eq.~(\ref{condition one driven}) with $r$, keeping all the other intrinsic parameters of the system as constant. The figure corresponds to a particular choice of the parameters $r_{wx}$, $r_{wy}$, $\lambda_c$ and $\lambda_d$. If the ratio $\Delta E_{tot}/\Delta E_{dif}$ which depends on the initial conditions of $P$ and $Q$, falls in the region below (above) the line in the phase diagram [see Fig.~\ref{fig01}(b)], then the system exhibits (does not exhibit) the Mpemba effect. 

For steady state initial conditions, the ratio $\Delta E_{tot}/\Delta E_{dif}$ is given by Eq.~(\ref{appendix: ss ratio Delta E}). As the ratio $\Delta E_{tot}/\Delta E_{dif}$ is a function of the driving strengths, $\sigma^2_x$ and $\sigma^2_y$ [see Eq.~(\ref{appendix: ss ratio Delta E})], they can be appropriately tuned, independently for the systems $P$ and $Q$ as well as along the $x$ and $y$ directions, to access the entire region of phase space where the Mpemba effect is observable.

Note that one can introduce anisotropy in the mean kinetic energies by simply considering the case $\sigma^2_x\neq \sigma^2_y$, and keeping $r_{wx}=r_{wy}$ [see Eqs.~(\ref{energy-x-ss}) and (\ref{energy-y-ss})]. But in that case, the condition for the Mpemba effect reduces to $\Delta E_{tot}<0$ [see Eq.~(\ref{condition one driven})] which is not possible to realise as we have assumed $\Delta E_{tot}=E^P_{tot}(0)-E^Q_{tot}(0)>0$. Therefore, to demonstrate the Mpemba effect, we restrict ourselves to the case $r_{wx}\neq r_{wy}$.

\subsection{\label{sec4c-The inverse Mpemba effect}The inverse Mpemba effect}
\begin{figure}
\centering
\includegraphics[width=0.8\columnwidth]{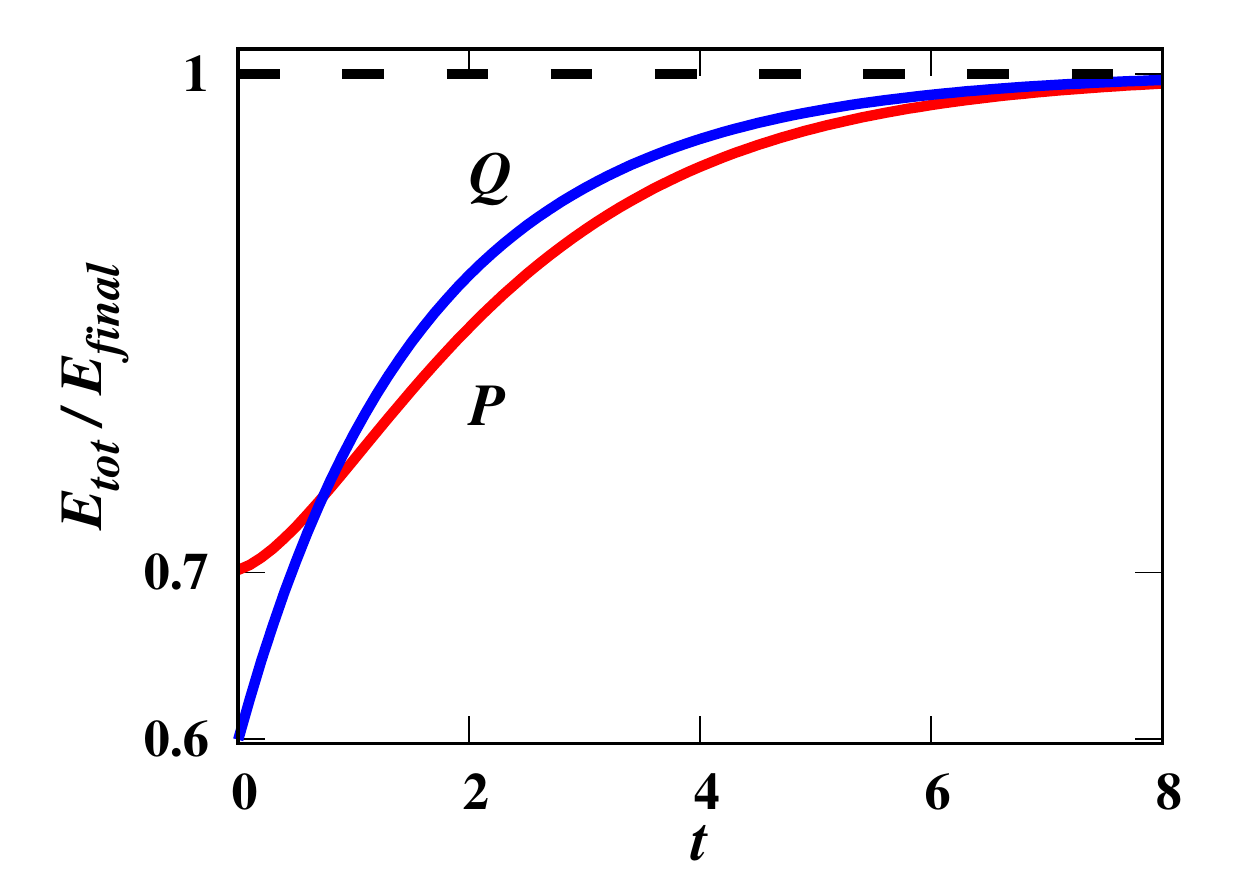}
\caption{The time evolution of the total energy, $E_{tot}(t)$ for anisotropically driven systems $P$ and $Q$ of a two dimensional inelastic Maxwell gas, driven along both the directions of the plane, with initial conditions $E^P_{tot}(0)$=20.27, $E^Q_{tot}(0)$=17.32, $E^P_{dif}(0)$=-7.93 and $E^Q_{dif}(0)$=6.26 such that $E^P_{tot}(0)>E^Q_{tot}(0)$, which satisfies the condition for the inverse Mpemba effect as described in Eq.~(\ref{condition one driven}). The other parameters decribing the systems are chosen to be $r$=0.3, $r_{wx}=0.88$ and $r_{wy}=0.39$. $P$ relaxes to the steady state slower than $Q$, though its initial energy is larger. The time at which the trajectories cross each other is $\tau=0.73$ as given by Eq.~(\ref{crossing time one driven}).}
\label{fig03}
\end{figure}

Consider now the case where a system is heated instead of being cooled unlike the direct Mpemba effect.  Now if an initially colder system heats up faster than a system at an intermediate one then it is called the inverse Mpemba effect. We follow the same analysis as in the direct Mpemba effect. The condition for the inverse Mpemba effect is same as that for the direct Mpemba effect as given in Eq.~(\ref{condition one driven}).  We prepare two systems $P$ and $Q$ such that $P$ has a higher initial total energy than $Q$ and also satisfy the condition for the inverse Mpemba effect [Eq.~(\ref{condition one driven})]. We then quench both the systems to a common steady state having higher total energy compared to the initial total energies of both $P$ and $Q$. The cross-over time $\tau$ at which the trajectories $E^P_{tot}(t)$ and $E^Q_{tot}(t)$ cross is given by Eq.~(\ref{crossing time one driven}). An example is illustrated in Fig.~\ref{fig03}.

The phase space of the initial steady states that satisfy the condition for the inverse Mpemba effect turns out to be the same as that for the direct Mpemba effect and is given by Eq.~(\ref{appendix: ss ratio Delta E}). Thus, Fig.~\ref{fig01}(b) also illustrates the valid initial steady states given by Eq.~(\ref{appendix: ss ratio Delta E}) that satisfy the condition [Eq.~(\ref{condition one driven})] where the inverse Mpemba effect is observable.

\subsection{\label{sec4d-The strong Mpemba effect}The strong Mpemba effect}
It can be shown that there exists certain initial conditions such that the system at higher temperature relaxes to a final steady state exponentially faster compared to other initial conditions. This phenomenon is called the strong Mpemba effect. The effect may be realised when the coefficient ($K_-$) associated with the slower relaxation rate  in the time evolution of total kinetic energy, $E_{tot}(t)$ [see Eq.~(\ref{sol E1 one driven})] vanishes.

Setting the coefficient $K_-$ [given by Eq.~(\ref{appendix coeff both components driven})]  to zero, we obtain
\begin{align}
E_{tot}(0)=\frac{2\lambda_d (r^2_{wy}-r^2_{wx})}{\lambda_c\alpha^2 + \sqrt{4 \lambda^2_d(r^2_{wy}-r^2_{wx})^2+ \alpha^4 \lambda^2_c}} E_{dif}(0)+c, \label{strong mpemba}
\end{align}

where
\begin{align}
c=\frac{2\lambda_d \left[2(\sigma^2_x + \sigma^2_y)  + \lambda_d (\sigma^2_x - \sigma^2_y) (r^2_{wx}-r^2_{wy})   \right]}{2 \lambda_d (2-r^2_{wx}-r^2_{wy}) + \alpha \lambda_c(4 -3 \alpha) - \sqrt{4 \lambda^2_d(r^2_{wy}-r^2_{wx})^2+ \alpha^4 \lambda^2_c}}.
\end{align}

The solution of Eq.~(\ref{strong mpemba}) in terms of 
$E_{tot}(0)$ and $E_{dif}(0)$  yields the set of initial states whose relaxation is 
exponentially faster than the  set of generic states.  Among these initial states one would like to determine the ones which are steady states. The steady state ratio of $E_{tot}(0)/E_{dif}(0)$~[see Eq.~(\ref{appendix coeff both components driven})] for a system is given by
\begin{align}
\frac{E_{tot}(0)}{E_{dif}(0)}=\frac{\begin{aligned} 
&\big[\lambda_d (1-r^2_{wy}) + \alpha \lambda_c(2-\alpha)\big] \sigma^2_x + \big[2\lambda_d (1-r^2_{wx}) + \alpha \lambda_c(2-\alpha)\big] \sigma^2_y
\end{aligned}}{\begin{aligned} 
&2\big[\lambda_d (1-r^2_{wy}) + \alpha \lambda_c(1-\alpha)\big] \sigma^2_x - \big[\lambda_d (1-r^2_{wx}) + \alpha \lambda_c(1-\alpha)\big] \sigma^2_y
\end{aligned}},
\end{align}
and is a function of the driving strengths, $\sigma^2_x$ and $\sigma^2_y$, as all other parameters are kept constant. 
One observes that the valid steady states with initial energies, $E_{tot}(0)$ and $E_{dif}(0)$ that satisfy the condition for the strong Mpemba effect [see Eq.~(\ref{strong mpemba})] can be obtained by appropriately tuning the driving strengths.

\begin{figure}
\centering
\includegraphics[width=0.8 \columnwidth]{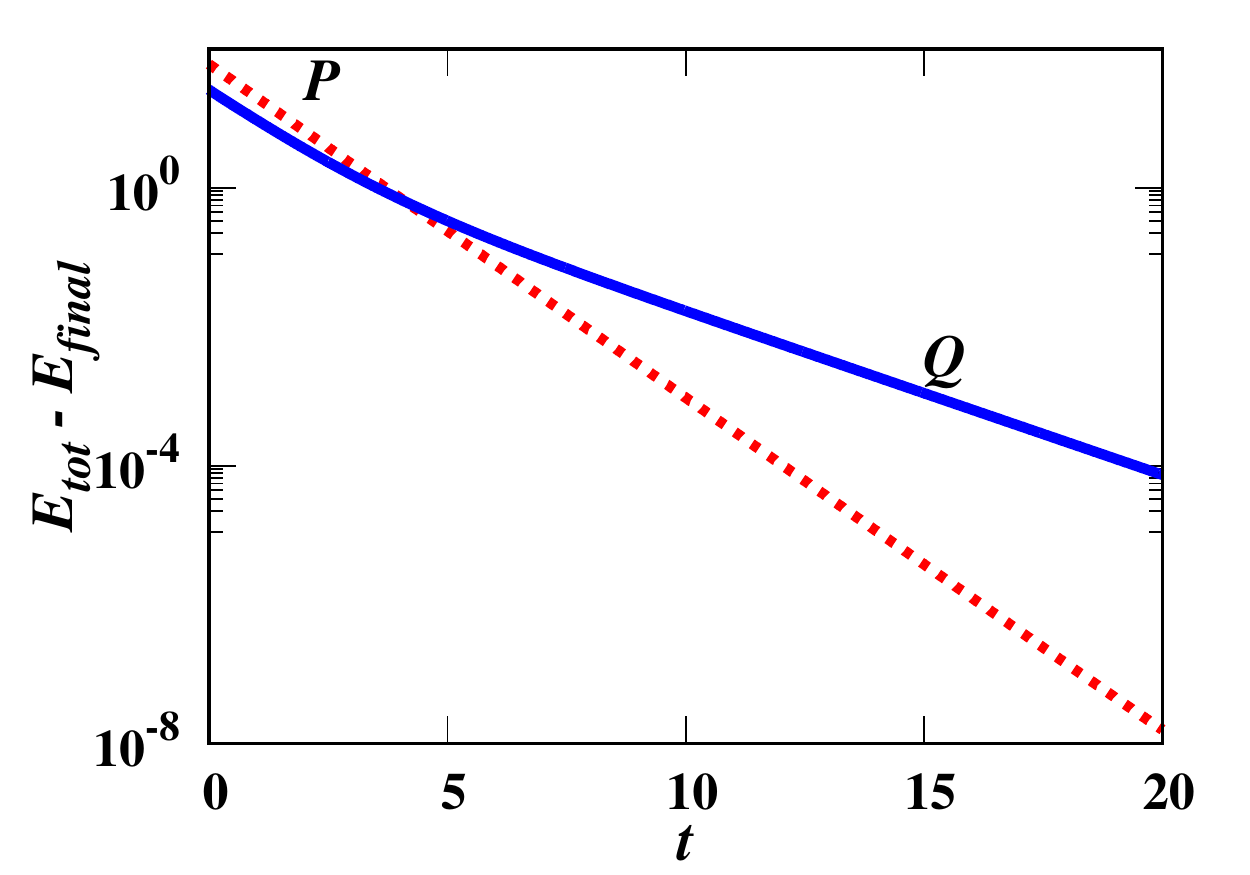}
\caption{The time evolution of the total energy, $E_{tot}(t)$ for anisotropically driven systems $P$ and $Q$ of a two dimensional inelastic Maxwell gas, driven along both the directions of the plane, with  $E^P_{tot}(0)=95.89$, $E^Q_{tot}(0)=61.57$, $E^P_{diff}(0)=-59.82$ and $E^Q_{diff}(0)=-4.26$ such that $E^P_{tot}(0)>E^Q_{tot}(0)$. These initial values satisfy both  the conditions for  the Mpemba effect as described in Eq.~(\ref{condition one driven}) as well as those  for   the strong Mpemba effect (for system $P$) as described in Eq.~(\ref{strong mpemba}). The other parameters defining the system are chosen to be $r=0.2$, $r_{wx}=0.88$ and $r_{wy}=0.49$. $P$ equilibrates  to the final state at an exponentially faster rate compared to $Q$ and the time at which the trajectories cross each other is $\tau=4.14$ as given by Eq.~(\ref{crossing time one driven}).}
\label{fig04}
\end{figure}

Thus, for a system of monodispersed Maxwell gas in two dimensions, there exists steady state initial conditions
that satisfy the condition  given by Eq.~(\ref{strong mpemba}) and hence approach the final steady state exponentially faster compared to  any other  similar system whose initial energies lie slightly below or above the line. An example of the strong Mpemba effect is shown  in Fig.~\ref{fig04}.

\section{\label{sec5-Limiting case where the driving is along one direction}Special case when the driving is only in one direction}
In Sec.~\ref{sec4-The Mpemba effect}, we discussed the possibility of the Mpemba effect in the case of anisotropically driven monodispersed Maxwell gas where the particles are driven along both the directions.  We now consider a similar system but the driving is restricted to one direction. We follow the same analysis as in Sec.~\ref{sec4-The Mpemba effect}. Here, for the case when particles are driven only along $x$-direction (say) with driving strengths, $\sigma^2_x\neq0$ and $\sigma^2_y=0$, the time evolution of mean kinetic energies $E_x$ and $E_y$ is
\begin{align}
\begin{split}
&\frac{dE_x(t)}{dt}=E_x \big[ \lambda_c\alpha(\frac{3}{4}\alpha -1) -\lambda_d(1-r^2_{wx}) \big] + E_y \big[ \frac{\lambda_c}{4}\alpha^2 \big] + \lambda_d \sigma^2_x, \\
&\frac{dE_y(t)}{dt}=E_x \big[ \frac{\lambda_c}{4}\alpha^2 \big] +E_y \big[ \lambda_c\alpha(\frac{3}{4}\alpha -1)  \big].
\end{split}
\end{align}
The time evolution for the quantities $E_{tot}$ and $E_{dif}$  are given by Eq.~(\ref{time ev E}) but now the column matrix $\boldsymbol{D}$ takes the form
\begin{align}
\boldsymbol{D}&=\begin{bmatrix}
\lambda_d \sigma^2_x, 
\lambda_d \sigma^2_x
\end{bmatrix}^T,
\end{align}
The solutions for $E_{tot}(t)$ and $E_{dif}(t)$ are obtained in the similar way as in Eq.~(\ref{sol E1 one driven}) with the coefficients $K_+, K_-, L_+$ and $L_-$ along with the steady state energies $\langle E_{tot}\rangle$ and $\langle E_{dif}\rangle$ given in Eq.~(\ref{appendix coeff one component driven}).

We now consider two  systems labeled as $P$ and $Q$ with different initial conditions $[E^P_{tot}(0), E^P_{dif}(0)]$ and $[E^Q_{tot}(0), E^Q_{dif}(0)]$ where $E^P_{tot}(0)> E^Q_{tot}(0)$. Both the systems are quenched to a  common steady state whose total energy is smaller than the initial total energies of $P$ and $Q$. This is achieved by changing the driving strengths of $P$ and $Q$ to the common driving strengths, $\sigma^2_x\neq0$ and $\sigma^2_y=0$ of the final steady state, keeping all the other parameters constant for both the systems.

\begin{figure}
\centering
\includegraphics[width= \columnwidth]{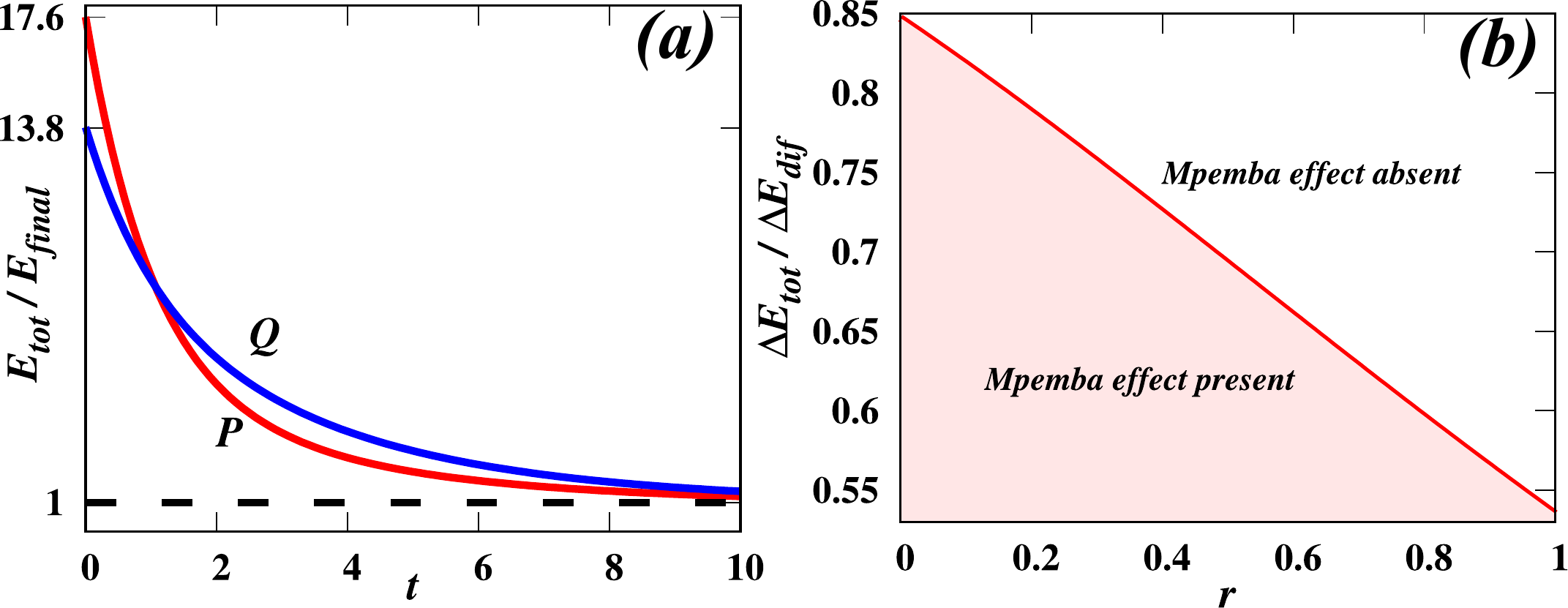}
\caption{(a) The time evolution of the total energy, $E_{tot}(t)$ for anisotropically driven systems $P$ and $Q$ of a two dimensional inelastic Maxwell gas driven along a single direction   with initial conditions $E^P_{tot}(0)$=28, $E^Q_{tot}(0)$=22, $E^P_{dif}(0)$=26 and $E^Q_{dif}(0)$=5 such that $E^P_{tot}(0)>E^Q_{tot}(0)$, which satisfies the condition for the Mpemba effect as described in Eq.~(\ref{condition one driven}). The other parameters decribing the systems are chosen to be $r$=0.5 and $r_{wx}=0.6$. $P$ relaxes to the steady state faster than $Q$, though its initial energy is larger. The time at which the trajectories cross each other is $\tau=1.07$ as given by Eq.~(\ref{crossing time one driven}). (b) $\Delta E_{tot}/\Delta E_{dif}$--$r$ phase diagram showing regions where the Mpemba effect is observed and $r$ is the coefficient of restitution. The line corresponds to a particular choice of the parameters $r_{wx}=0.5$, $\lambda_c=1.0$ and $\lambda_d=1.0$. The region below the line given by Eq.~(\ref{condition one driven}) shows the Mpemba effect whereas the region on the other side of the line does not show the Mpemba effect.}
\label{fig05}
\end{figure}

The condition for the Mpemba effect to be present is the same as that derived for the more general case [see Eq.~(\ref{condition one driven})]. In Fig.~\ref{fig05}(a), we consider such a situation where Eq.~(\ref{condition one driven}) is satisfied and hence the systems $P$ and $Q$ show the Mpemba effect. The trajectories cross at the point as predicted by Eq.~(\ref{crossing time one driven}).

In Fig.~\ref{fig05}(b), we identify the region of phase space (initial condition) where the Mpemba effect is observable, based on Eq.~(\ref{condition one driven}). In the figure, the line denotes the variation of right hand side of Eq.~(\ref{condition one driven}) with $r$, keeping all the other intrinsic parameters of the system as constant. The figure corresponds to a particular choice of the parameters $r_{wx}$, $\lambda_c$ and $\lambda_d$. The region below the line in the phase diagram corresponds to the initial conditions $\Delta E_{tot}/\Delta E_{dif}$ [see Fig.~\ref{fig05}(b)]  that show the Mpemba effect [Eq.~(\ref{condition one driven})] whereas the other region does not show the effect.

Here, we consider that the systems $P$ and $Q$ have identical intrinsic parameters once the quench is done to the common steady state. However, these intrinsic parameters that characterise the initial conditions of $P$ and $Q$ or equivalently $\Delta E_{tot}/\Delta E_{dif}$,  could be different.  As a result, one can tune these intrinsic parameters differently for $P$ and $Q$ to obtain initial steady states that satisfy the condition given by Eq.~(\ref{condition one driven}) and hence show the Mpemba effect.

However, when the intrinsic parameters other than driving strength is kept the same (both before and after the quench), the ratio $\Delta E_{tot}/\Delta E_{dif}$  for  initial steady states has a simple form:
\begin{equation}
\frac{\Delta E_{tot}}{\Delta E_{dif}}=\frac{(2-\alpha)}{2(1-\alpha)} \ge 1.5. \label{steady state condition mpemba}
\end{equation}

Note that $ \alpha \in[1/2, 1]$ and hence the ratio in Eq.~(\ref{steady state condition mpemba}) is always larger than or equal to 1.5. However, we know from Eq.~(\ref{condition one driven}) that for the Mpemba effect to be present, $\Delta E_{tot}/\Delta E_{dif}<1$. Thus, Eq.~(\ref{steady state condition mpemba}) does not satisfy the required condition for the existence of the Mpemba effect. We conclude that for initial states that correspond to steady states where $P$ and $Q$ have identical intrinsic parameters except for the driving strength, the Mpemba effect is not possible when the driving is restricted to one direction.

\section{\label{sec6-Summary and discussion}Summary and discussion}
In this paper, we have shown an exact analysis for the existence of the Mpemba effect, the inverse Mpemba effect and the strong Mpemba effect in an anisotropically driven inelastic Maxwell gas in two dimensions. The Maxwell model for granular gases is a simplified model where the rate of collision between the granular particles is independent of their relative velocities. In addition, we assumed the well-mixed limit such that the spatial correlations were ignored. The model allows for an exact solution as the two-point  velocity correlations form a coupled set of linear equations.

We show that anisotropic driving leads to the existence of the Mpemba effect starting from steady state initial conditions unlike the case of isotropic driving which required the initial conditions to be non-stationary. We considered two different cases of anisotropic driving in two dimensions: when particles are driven along one direction only and the other case where particles are driven along both the directions. In both the cases, we show that the Mpemba effect can exist for initial conditions which are valid steady states characterised by the parameters of the system. We also demonstrated the existence of the inverse Mpemba effect where a system is heated instead of being cooled. Here, an initially colder system equilibrates to a final high temperature state faster than an initially warmer system. We also derived the condition for the existence of the strong Mpemba effect where for certain initial states, a system equilibrates at an exponentially faster rate compared to any other states.

Our results place the results obtained by us in an earlier work~\cite{biswas2021mpemba} on an anisotropically driven granular gas with a more realistic velocity dependent collision rate, on a more rigorous footing. One difference between results for the two models is that in the Maxwell gas, the results depend on whether the anisotropy is applied through different driving strengths ($\sigma^2_x\neq \sigma^2_y$) or through different driving parameters ($r_{wx}\neq r_{wy}$). For the Maxwell gas, if all other parameters are kept the same, then the latter condition is necessary. This difference is due to the lack of non-linear coupling between $E_{tot}$ and $E_{dif}$ in the Maxwell model.

\begin{appendices}
\appendix

\section{\label{ap: Time evolutions for mean kinetic energies}Coefficients for the time evolutions of $E_{tot}$ and $E_{dif}$}
In this Appendix, we solve for the time evolutions of $E_{tot}$ and $E_{dif}$ for the anisotropically driven inelastic Maxwell gas in two dimensions. We consider two cases of anisotropic driving: when the external driving is applied along both the directions with different driving strengths and another case where the driving is along one direction only as described in \ref{ap: driving is along both directions} and \ref{ap: driving is along single direction} respectively.

\subsection{\label{ap: driving is along both directions}Driven along both directions of the plane}
We consider an inelastic Maxwell gas in two dimensions where the particles are driven along both directions of the plane at rate $\lambda_d$ and with different driving strengths, $\sigma^2_x$ and $\sigma^2_y$ respectively. The time evolutions of $E_{tot}$ and $E_{dif}$ are given as in Eq.~(\ref{sol E1 one driven}) where the coefficients $K_+, K_-, L_+$ and $L_-$ along with the steady state energies $\langle E_{tot}\rangle$ and $\langle E_{dif}\rangle$ are given by
 {\footnotesize
\begin{align}
K_+&=\frac{1}{\gamma}\Big[ -(\lambda_--\chi_{11})E_{tot}(0)+\chi_{12}E_{dif}(0)-\frac{\lambda_d}{\lambda_+}\big[\big( \chi_{12}-(\lambda_--\chi_{11})  \big)\sigma^2_x - \big( \chi_{12}+(\lambda_--\chi_{11})  \big) \sigma^2_y \big] \Big], \nonumber\\
K_-&= \frac{1}{\gamma}\Big[ (\lambda_+-\chi_{11})E_{tot}(0)-\chi_{12}E_{dif}(0)+ \frac{\lambda_d}{\lambda_-}\big[\big( \chi_{12}-(\lambda_+-\chi_{11})  \big)\sigma^2_x - \big( \chi_{12}+(\lambda_+-\chi_{11})  \big) \sigma^2_y \big] \Big],\nonumber \\
\langle E_{tot}\rangle&=\frac{\lambda_d}{\gamma}\Big[ \frac{\big( \chi_{12}-(\lambda_--\chi_{11})  \big)\sigma^2_x - \big( \chi_{12}+(\lambda_--\chi_{11})  \big) \sigma^2_y}{\lambda_+} - \frac{\big( \chi_{12}-(\lambda_+-\chi_{11})  \big)\sigma^2_x - \big( \chi_{12}+(\lambda_+-\chi_{11})  \big) \sigma^2_y}{\lambda_-} \Big],\nonumber\\
L_+&=\frac{1}{\gamma}\Big[ -\frac{(\lambda_+-\chi_{11})(\lambda_--\chi_{11})}{\chi_{12}}E_{tot}(0) + (\lambda_+ - \chi_{11})E_{dif}(0) -\frac{\lambda_d}{\lambda_+ \chi_{12}}\big[(\lambda_+-\chi_{11})(\lambda_--\chi_{11}) (\sigma^2_x- \sigma^2_y) \big] \Big], \nonumber\\ 
L_-&=\frac{1}{\gamma}\Big[ \frac{(\lambda_+-\chi_{11})(\lambda_--\chi_{11})}{\chi_{12}}E_{tot}(0) - (\lambda_+ - \chi_{11})E_{dif}(0) +\frac{\lambda_d}{\lambda_- \chi_{12}}\big[(\lambda_+-\chi_{11})(\lambda_--\chi_{11}) (\sigma^2_x- \sigma^2_y) \big] \Big], \nonumber\\ 
\langle E_{dif}\rangle&=\frac{\lambda_d}{\chi_{12}\gamma} \Big[ (\lambda_+-\chi_{11})(\lambda_--\chi_{11}) (\sigma^2_x- \sigma^2_y) \big( \frac{1}{\lambda_+}-\frac{1}{\lambda_-} \big) \Big]            , \nonumber\\ 
\gamma&=\lambda_+-\lambda_-.    \label{appendix coeff both components driven}
\end{align}
}

\subsection{\label{ap: driving is along single direction}Driven along a single direction of the plane}
Here, we consider an inelastic Maxwell gas in two dimensions where the particles are driven along a single direction (say along $x$ direction) at a rate $\lambda_d$ and  with driving strengths, $\sigma^2_x\neq0$ and $\sigma^2_y=0$.
The time evolutions of $E_{tot}$ and $E_{dif}$ are given as in Eq.~(\ref{sol E1 one driven}) where the coefficients $K_+, K_-, L_+$ and $L_-$ along with the steady state energies $\langle E_{tot}\rangle$ and $\langle E_{dif}\rangle$ are now given by 
{\footnotesize
\begin{align}
\begin{split}
K_+&=\frac{1}{\gamma}\Big[ (-\lambda_-+\chi_{11})E_{tot}(0)+\chi_{12}E_{dif}(0)-\frac{\chi_{12}-\lambda_-+\chi_{11})}{\lambda_+}\lambda_d \sigma^2_x \Big], \\
K_- &= \frac{1}{\gamma}\Big[ (\lambda_+-\chi_{11})E_{tot}(0)-\chi_{12} E_{dif}(0)+\frac{\chi_{12}-\lambda_++\chi_{11})}{\lambda_-}\lambda_d \sigma^2_x \Big], \\
\langle E_{tot}\rangle&=\frac{1}{\gamma}\Big[ \frac{\chi_{12}-(\lambda_--\chi_{11})}{\lambda_+}-\frac{\chi_{12}-(\lambda_+ -\chi_{11})}{\lambda_-}   \Big]\lambda_d \sigma^2_x,\\
\langle E_{dif}\rangle&=\frac{1}{\gamma}\Big[ \frac{(\lambda_+-\chi_{11})(\lambda_--\chi_{11})}{\chi_{12}\lambda_+}- \frac{(\lambda_+-\chi_{11})(\lambda_--\chi_{11})}{\chi_{12}\lambda_-}  \Big]\lambda_d \sigma^2_x, \\
L_+&=\frac{1}{\gamma}\Big[ -\frac{(\lambda_+-\chi_{11})(\lambda_--\chi_{11})}{\chi_{12}}E_{tot}(0) + (\lambda_+ - \chi_{11})E_{dif}(0) -\frac{(\lambda_+-\chi_{11})(\lambda_--\chi_{11})}{\chi_{12}\lambda_+}\lambda_d \sigma^2_x \Big], \\
L_-&=\frac{1}{\gamma}\Big[ \frac{(\lambda_+-\chi_{11})(\lambda_--\chi_{11})}{\chi_{12}}E_{tot}(0) - (\lambda_- - \chi_{11})E_{dif}(0) +\frac{(\lambda_+-\chi_{11})(\lambda_--\chi_{11})}{\chi_{12}\lambda_-}\lambda_d \sigma^2_x \Big],\\
\gamma&=\lambda_+-\lambda_-. 
\end{split}\label{appendix coeff one component driven}
\end{align}
}

\end{appendices}

\bibliographystyle{iopart-num}

\begin{thebibliography}{10}
\expandafter\ifx\csname url\endcsname\relax
  \def\url#1{{\tt #1}}\fi
\expandafter\ifx\csname urlprefix\endcsname\relax\def\urlprefix{URL }\fi
\providecommand{\eprint}[2][]{\url{#2}}

\bibitem{1952meteorologica}
Lee H {\em Meteorologica,\/} Loeb classical library: Greek authors p.{\bf 87}
  (Harvard University Press, Cambridge, MA, 1952)

\bibitem{Mpemba_1969}
Mpemba E~B and Osborne D~G 1969 {\em Phys. Educat.\/} {\bf 4} 172--175

\bibitem{vynnycky-convection:2015}
Vynnycky M and Kimura S 2015 {\em Int. J. Heat Mass Transf.\/} {\bf 80} 243 --
  255

\bibitem{Mirabedin-evporation-2017}
Mirabedin S~M and Farhadi F 2017 {\em Int. J. Refrig.\/} {\bf 73} 219 -- 225

\bibitem{katz2009hot}
Katz J~I 2009 {\em Am. J. Phys.\/} {\bf 77} 27--29

\bibitem{david-super-cooling-1995}
Auerbach D 1995 {\em Am. J. Phys.\/} {\bf 63} 882--885

\bibitem{zhang-hydrbond1-2014}
Zhang X, Huang Y, Ma Z, Zhou Y, Zhou J, Zheng W, Jiang Q and Sun C~Q 2014 {\em
  Phys. Chem. Chem. Phys.\/} {\bf 16}(42) 22995--23002

\bibitem{tao-hydrogen-2017}
Tao Y, Zou W, Jia J, Li W and Cremer D 2017 {\em J. Chem. Theory Comput.\/}
  {\bf 13} 55--76

\bibitem{Molecular_Dynamics_jin2015mechanisms}
Jin J and Goddard~III W~A 2015 {\em J. Phys. Chem. C\/} {\bf 119} 2622--2629

\bibitem{gijon2019paths}
Gij{\'o}n A, Lasanta A and Hern{\'a}ndez E 2019 {\em Phys. Rev. E\/} {\bf 100}
  032103

\bibitem{NoMpemba}
Burridge H~C and Linden P~F 2016 {\em Sci. Rep.\/} {\bf 6} 37665

\bibitem{paper:hydrates}
Ahn Y~H, Kang H, Koh D~Y and Lee H 2016 {\em Korean J. Chem. Eng.\/} {\bf 33}
  1903--1907

\bibitem{chaddah2010overtaking}
Chaddah P, Dash S, Kumar K and Banerjee A 2010 {\em arXiv preprint
  arXiv:1011.3598\/}

\bibitem{Polylactide}
Hu C, Li J, Huang S, Li H, Luo C, Chen J, Jiang S and An L 2018 {\em Cryst.
  Growth Des.\/} {\bf 18} 5757--5762

\bibitem{kumar2020exponentially}
Kumar A and Bechhoefer J 2020 {\em Nature\/} {\bf 584} 64--68

\bibitem{kumar2021anomalous}
Kumar A, Chetrite R and Bechhoefer J 2021 {\em arXiv preprint
  arXiv:2104.12899\/}

\bibitem{PhysRevLett.124.060602}
Gal A and Raz O 2020 {\em Phys. Rev. Lett.\/} {\bf 124}(6) 060602

\bibitem{Klich-2019}
Klich I, Raz O, Hirschberg O and Vucelja M 2019 {\em Phys. Rev. X\/} {\bf 9}(2)
  021060

\bibitem{klich2018solution}
Klich I and Vucelja M 2018 {\em arXiv preprint arXiv:1812.11962\/}

\bibitem{das2021should}
Das S~K and Vadakkayil N 2021 {\em Phys. Chem. Chem. Phys.\/}

\bibitem{Lu-raz:2017}
Lu Z and Raz O 2017 {\em Proc. Natl. Acad. Sci. USA\/} {\bf 114} 5083--5088

\bibitem{SpinGlassMpemba}
Baity-Jesi M, Calore E, Cruz A, Fernandez L~A, Gil-Narvi{\'o}n J~M,
  Gordillo-Guerrero A, I{\~n}iguez D, Lasanta A, Maiorano A, Marinari E {\em
  et~al.\/} 2019 {\em Proc. Natl. Acad. Sci. USA\/} {\bf 116} 15350--15355

\bibitem{moleculargas}
Santos A and Prados A 2020 {\em Phys. Fluids\/} {\bf 32} 072010

\bibitem{gonzalez2020mpemba}
Gonz{\'a}lez R~G, Khalil N and Garz{\'o} V 2020 {\em arXiv preprint
  arXiv:2010.14215\/}

\bibitem{gonzalez2020anomalous}
Gonz{\'a}lez R~G and Garz{\'o} V 2020 {\em arXiv preprint arXiv:2011.13237\/}

\bibitem{Lasanta-mpemba-1-2017}
Lasanta A, Vega~Reyes F, Prados A and Santos A 2017 {\em Phys. Rev. Lett.\/}
  {\bf 119}(14) 148001

\bibitem{Torrente-rough-2019}
Torrente A, L\'opez-Casta\~no M~A, Lasanta A, Reyes F~V, Prados A and Santos A
  2019 {\em Phys. Rev. E\/} {\bf 99}(6) 060901

\bibitem{mompo2020memory}
Momp{\'o} E, Casta{\~n}o M, Torrente A, Reyes F~V and Lasanta A 2020 {\em arXiv
  preprint arXiv:2006.00241\/}

\bibitem{PhysRevE.102.012906}
Biswas A, Prasad V~V, Raz O and Rajesh R 2020 {\em Phys. Rev. E\/} {\bf 102}(1)
  012906

\bibitem{biswas2021mpemba}
Biswas A, Prasad V~V and Rajesh R 2021 {\em arXiv preprint arXiv:2104.08730\/}

\bibitem{PhysRevE.103.032901}
Takada S, Hayakawa H and Santos A 2021 {\em Phys. Rev. E\/} {\bf 103}(3) 032901

\bibitem{PhysRevE.66.011309}
Ben-Naim E and Krapivsky P~L 2002 {\em Phys. Rev. E\/} {\bf 66}(1) 011309

\bibitem{Baldassarri_2002}
Baldassarri A, Marconi U~M~B and Puglisi A 2002 {\em Europhys. Lett.\/} {\bf
  58} 14--20

\bibitem{Ernst_2002}
Ernst M~H and Brito R 2002 {\em Europhys. Lett.\/} {\bf 58} 182--187

\bibitem{Krapivsky_2002}
Krapivsky P~L and Ben-Naim E 2002 {\em J. Phys. A\/} {\bf 35} L147--L152

\bibitem{PhysRevE.65.040301}
Ernst M~H and Brito R 2002 {\em Phys. Rev. E\/} {\bf 65}(4) 040301

\bibitem{PhysRevE.66.062301}
Antal T, Droz M and Lipowski A 2002 {\em Phys. Rev. E\/} {\bf 66}(6) 062301

\bibitem{PhysRevE.68.011305}
Santos A and Ernst M~H 2003 {\em Phys. Rev. E\/} {\bf 68}(1) 011305

\bibitem{prasad2014high}
Prasad V~V, Sabhapandit S and Dhar A 2014 {\em Europhys. Lett.\/} {\bf 104}
  54003

\bibitem{Prasad:14}
Prasad V~V, Sabhapandit S and Dhar A 2014 {\em Phys. Rev. E\/} {\bf 90}(6)
  062130

\bibitem{prasad2017velocity}
Prasad V~V, Das D, Sabhapandit S and Rajesh R 2017 {\em Phys. Rev. E\/} {\bf
  95} 032909

\bibitem{binary_maxwell}
Biswas A, Prasad V~V and Rajesh R 2020 {\em J. Stat. Mech.: Theory Exp.\/} {\bf
  2020}(1) 013202

\bibitem{Prasad:19}
Prasad V~V, Das D, Sabhapandit S and Rajesh R 2019  {\em J. Stat. Mech.: Theory
  Exp.\/} {\bf 2019}(6) 063201

\bibitem{Prasad:18}
Prasad V~V and Rajesh R 2019 {\em J. Stat. Phys.\/} {\bf 176} 1409–1433

\end{thebibliography}
\providecommand{\newblock}{}

\end{document}